# Core energy and regularization parameters of non-singular continuum theories of dislocations using atomistic simulations

Kamyar M Davoudi[1]

Department of Chemical and Materials Engineering, Faculty of Engineering, University of Alberta, Edmonton, AB, Canada
*Current Address:* Avidemia Education Inc, North Vancouver, BC, Canada

## Abstract

The dislocation core is an important region as it controls many important properties of materials. Elasticity breaks down in the core and the stress, force, and energy diverge at the dislocation line. We consider three commonest methods employed in Discrete Dislocation Dynamics (DDD) simulations to eliminate these singularities: (1) considering a cutoff parameter, (2) spreading the Burgers vector (CAWB theory), and (3) using gradient elasticity. Each of these methods includes an extra length parameter to regularize the elastic fields. In this article, we show that these regularization parameters can significantly influence the results of the DDD simulations. We use atomistic simulations for mixed dislocations to find the radius and energy of the dislocation core and find the regularization parameter and its variations with the dislocation character angle in each of the three methods. We have also considered if an arbitrary constant is chosen for the regularization parameter how the core energy should be added to the simulation codes. We have concluded that while the core energy in classical elasticity with a cutoff parameter can be described by one parameter, the other two methods need two energy parameters (core energy of edge and core energy of screw) for describing the variation of the core energy with the character angle. We have shown that no regularization parameter can be selected for the CAWB theory or gradient elasticity if no core energy is included.

## 1. Introduction

The dislocation core has two different meanings in the literature: the physical core and the mathematical core. The physical core is a severely distorted region around the dislocation line in which the coordination number and inversion symmetry of atoms are drastically different from that in a perfect crystal [1]. Dislocation core is an important region as its structure can significantly influence the behavior of dislocations and thereby its properties [2]. For example, the core structure of screw dislocations in body-centered cubic (bcc) metals influence the Peierls stress, the mobility of these dislocations, and therefore the low-temperature plastic deformation in these metals [3–5]. The formation energy of this distorted region in a perfect crystal is the core energy [6]. The core radius and the energy can be extracted from *ab initio* or atomistic calculations. In this article, the radius and the energy of the physical core are denoted by $r_{\text{core}}$ and $E_{\text{core}}$, respectively.

---
[1] Email: kamyar@avidemia.com, davoudi@alumni.harvard.edu

The distortion in the dislocation core is too large to be described by linear elasticity, and when classical elasticity is applied, the resulting stress, strain, and energy approach infinity as the distance from the dislocation approaches zero. These singularities are physically meaningless, and numerically very problematic [7–9]. Discrete Dislocation Dynamics (DDD) codes take various approaches to resolve this issue (see [10] and references therein). In the simplest approach, a cutoff parameter $\rho$ is introduced, and it is assumed that two dislocation segments do not interact with each other when they are closer than the cutoff parameter. This cutoff parameter is half of the core cutoff radius $r_0$ [11]. This is a mathematical core and its radius can be any arbitrary number as long as a core energy is added to compensate for the energy difference between the real energy of the crystal and what the model predicts. We denote the core radius and the core energy of this approach by $r_0$ and $E_c$, respectively. However, because the core energy is not introduced in many DDD codes, we have to choose $\rho$ (or $r_0$) such that we can ignore the core energy and let $E_c = 0$.

A different approach to the elimination of the singularities is spreading the Burgers vector. Peierls [12] and Nabarro [13] spread out the Burgers vector in the glide plane whereas Cai et al. [10] (hereafter referred to as the CAWB theory) spread it in all 3-dimensions about every point on the dislocation line. In the CAWB theory, the spreading function is characterized by a single parameter called the spreading radius and more often the core width or the core radius $r_c$ [8,10,14,15].

Using generalized continuum theories instead of the classical continuum theory is an alternative approach, which may lead to the elimination of the singularities from the stress field and/or the strain field [9]. In this article, we discuss two generalized theories, namely nonlocal elasticity [16,17] and strain gradient elasticity, in which one or two extra parameters with the dimensions of length are introduced in the constitutive laws [18–21]. In nonlocal elasticity, the stress at a point depends on the strain of the entire body; gradient elasticity has a weak nonlocality and stress is a function of the strain at the point and the nearby points (or the gradient of strain). For a comparison between nonlocal elasticity and gradient elasticity see, for example, Ref. [7,22].

No matter what theory we use to describe the elastic fields of a dislocation, we need a regularization parameter with the dimensions of length. In the classical theory of elasticity, we need to apply a cutoff parameter $\rho$, in the CAWB theory, we need to choose the core radius $r_c$, and in gradient elasticity, we need to specify the gradient coefficient $c$. Similar to what we mentioned about the cutoff parameter, $r_c$ and $c$ can be selected arbitrarily provided that a core energy is added to the elastic energy of the dislocation. If the core energy is ignored as in most of the DDD codes, the values of the regularization parameters must be selected such that the elastic energy of a dislocation in these theories equals the actual strain energy of a dislocation. Because the dependence of the strain energy of a dislocation on the dislocation character angle might be different from the dependence of elastic energy in these nonsingular theories, considering one energy parameter $E_c$ might be insufficient for this end.

In simulations some typical values are used for the regularization parameter and the core energy (if it is included). In section 2, we show that the effect of this regularization parameter is significant. In section 3, we use atomistic simulations to find the physical core radius $r_{\text{core}}$ and $E_{\text{core}}$. We perform the atomistic study for tungsten because (1) tungsten is a fairly good isotropic material, (2) it is a bcc material in which dislocations do not dissociate [23], and therefore we do not need to consider partial dislocations. In sections 4-6, for each of the nonsingular theories, we investigate if one energy parameter is sufficient to get an energy equal to the results of the atomistic calculations for all dislocation characters when the regularization parameter is chosen arbitrarily. Also, we investigate if the regularization parameter can be selected so that the core energy can be completely ignored.

## 2. The Importance of the Regularization Parameters

To investigate the role of the regularization parameters in nonsingular dislocation theories, we consider Frank-Read sources in tungsten with various initial lengths $L$ and calculate the nucleation stress. The nucleation stress of the Frank-Read sources directly impacts the stress-strain curve [24].

In the classical theory of elasticity, the nucleation stress $\sigma_{\text{nuc}}$ of a Frank-Read source of initial length $L$ is roughly given by [11]

$$\sigma_{\text{nuc}} = \frac{Gb}{2\pi L(1-\nu)} \left\{ \left[1 - \frac{\nu}{2}(3 - 4\cos^2\theta)\right] \ln\frac{L}{\rho} - 1 + \frac{\nu}{2} \right\} \tag{3}$$

where $G$ is the shear modulus, $b$ is the magnitude of the Burgers vector, $\nu$ is Poisson's ratio, and $\theta$ is the character angle formed between the line direction and the Burgers vector of the initial segment. The nucleation stress is calculated for some typical initial lengths of Frank-Read sources in thin films when $\theta = 0$ (Figure 1(a)) and when $\theta = 90°$ (Figure 1(b)). In thin films, Frank-Read sources are typically between 100$b$ and 1000$b$ long [25–29]. As we can see from Figure 1, the nucleation stress strongly depends on the cutoff parameter $\rho$. For example, when $L$ = 200$b$ and $\theta = 0$, the nucleation stress decreases from 630.67 MPa for $\rho = 0.2b$ to 394.62 MPa for $\rho = 2b$ (showing 59.8% reduction). More accurate calculations using some DDD codes that use a cutoff radius will show a similar pattern.

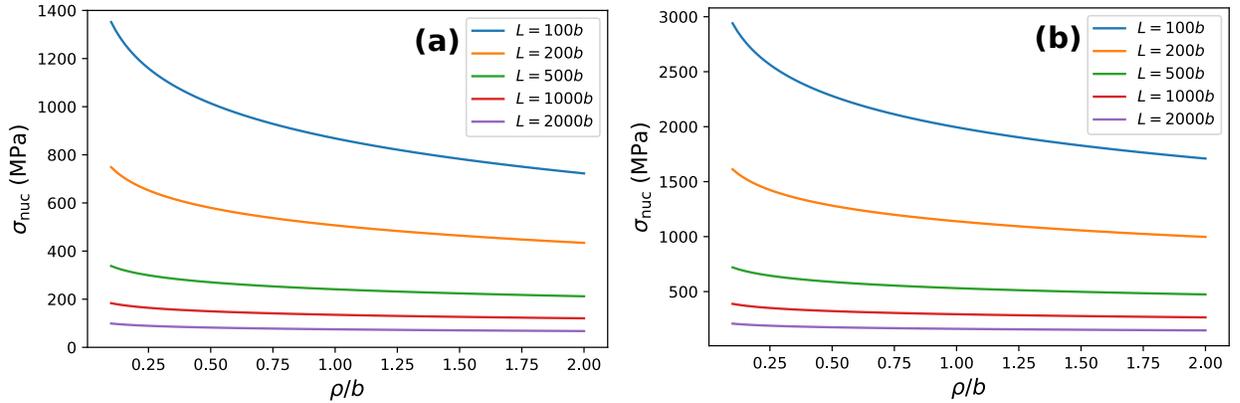

Figure 1: Nucleation stress of a Frank-Read source for various initial lengths $L$ estimated by Equation (3) when (a) $\theta$ = 90° and (b) $\theta$ = 0.

Discrete Dislocation Dynamics simulations have been performed using DDLab to study the effect of the core radius $r_c$ in the CAWB theory on the nucleation stress of an originally screw Frank-Read source (Figure 2). In these calculations, we have ignored the core energy and the Peierls stress to focus on the effect of the core radius alone. As we can see from Figure 2, the nucleation stress strongly depends on the core radius. For example, for the initial length of $L$ = 200$b$, when $r_c = 0.2b$ the nucleation stress is 1117.90 MPa and when $r_c = 2b$, the nucleation stress decreases to 709.91 MPa (showing 34.5% reduction).

Although we do not repeat the calculations for gradient elasticity, it is reasonably plausible that the effect of the gradient coefficient on the nucleation stress is extensive.

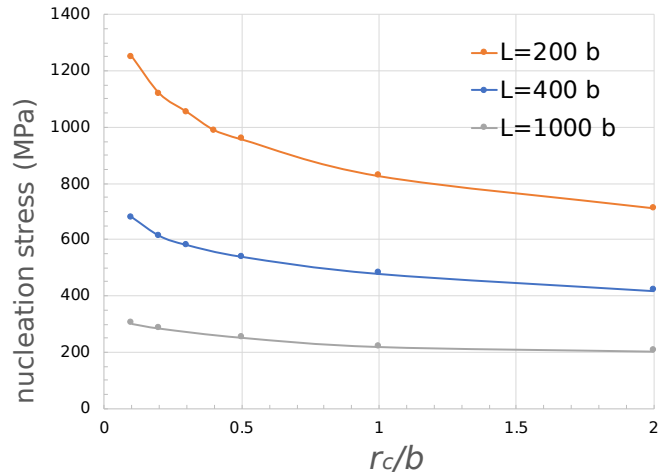

Figure 2: Nucleation stress of a frank-read source for different initial lengths *l* and various core width $r_c$ in the CAWB theory calculated by the ddd simulations.

## 3. Simulations

To investigate the core energy of mixed dislocations, a straight dislocation inside a slab was considered as shown in Figure 3, and periodic boundary conditions were imposed along the dislocation line (the *z* direction) and the possible gliding direction (the *x* direction). Because tungsten is a bcc material, the Burgers vector is $\mathbf{b} = a\langle 111 \rangle/2$ and the $\{111\}$ slip planes are preferred at low temperatures [2,3]. Table 1 shows how several mixed dislocations (with various character angles $\theta$ between the dislocation sense vector $\xi$ and the Burgers vector $\mathbf{b}$) were modeled by setting the *x, y,* and *z* directions.

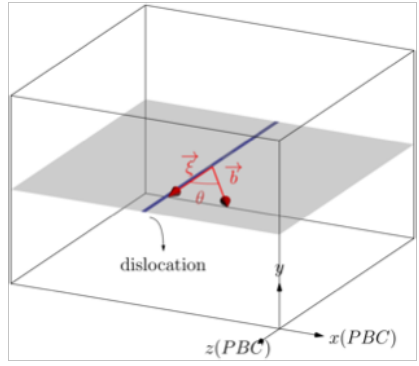

Figure 3: schematic modeling of a straight dislocation with an arbitrary character angle

| θ (degree) | x direction | y direction | z direction |
|---|---|---|---|
| 0 | $[\bar{1}2\bar{1}]$ | $[\bar{1}01]$ | $[111]$ |
| 8.05 | $[2\bar{3}2]$ | $[10\bar{1}]$ | $[343]$ |
| 19.47 | $[1\bar{1}1]$ | $[10\bar{1}]$ | $[121]$ |
| 35.26 | $[2\bar{1}2]$ | $[10\bar{1}]$ | $[141]$ |
| 44.71 | $[4\bar{1}4]$ | $[10\bar{1}]$ | $[181]$ |
| 51.06 | $[151]$ | $[\bar{1}01]$ | $[5\bar{2}5]$ |
| 70.53 | $[121]$ | $[\bar{1}01]$ | $[1\bar{1}1]$ |
| 90 | $[111]$ | $[\bar{1}01]$ | $[1\bar{2}1]$ |

Table 1: Simulation box directions for selected character angles.

A straight dislocation was inserted into the simulation box using the displacements predicted by the elastic solutions for a Volterra dislocation [30]. The uppermost and the lowermost layers of atoms in the *y* direction were fixed and the rest of the atoms were fully relaxed to their zero force positions at zero temperature [27,31]. Periodic boundaries along the *x* and *z* directions were recovered by removing a slab of length $b\sin\theta$ and titling the simulation box with respect to the *y* axis by $-0.5b\cos\theta$ [32,33]. The system was non-periodic along the *y* axis.

Because elastic constants directly impact the strain energy of dislocations, after testing several available potentials, the embedded atom method (EAM) potential developed by Han et al. [34] was used in this study. This potential produces the elastic constants that are close to the experimental results provided in Ref. [35] (see Table 2). All calculations were carried out using the parallel MD code LAMMPS, developed in Sandia National Lab [36] .

|  | Experiment | Simulation |
|---|---|---|
| Lattice constant (Å) | 3.1652 | 3.1592 |
| $C_{11}$ (GPa) | 532.55 | 532.61 |
| $C_{12}$ (GPa) | 204.95 | 205.02 |
| $C_{44}$ (GPa) | 163.13 | 163.20 |

Table 2: Comparison between lattice constant and elastic constants obtained in the experiments [32] and from the simulations using the EAM potential developed by Han et al. [31].

To make sure that the effect of the size of simulation box can be neglected, at least two different sample sizes for each character angles were studied. The smallest simulation box that was considered contained 614,952 atoms. Systems of quadrupole dislocations were also investigated to assure that the effect of the free surfaces is negligible.

Depending on the sign of the Burgers vector and the location of the dislocation, a screw dislocation in a bcc material can assume two different configurations: easy core and hard core. In the easy core structure, the distances between atoms in the core triangle is the same as that in the perfect crystal, whereas the hard core brings the triangle of atoms forming the dislocation to the same {111} plane and makes the distance between them much smaller than that in the perfect crystal [23,37,38]. Therefore, the hard core structure has a higher energy configuration than the easy core. It is often observed in simulations that the hard core structure is unstable or metastable [23,37,38]. The hard core structure could not be modeled as a single dislocation in the slab because upon revering the direction of the Burgers vector, the dislocation moves to the easy core configuration. However, both the easy core and the hard core structures were observed in the square-like periodic array of quadrupoles (also see [37]). For these reasons, the hard core structure is often considered as an energy barrier between two adjacent easy core structures [23], and in this article, we have considered only the energy of a screw dislocation with the easy core structure.

## 4. Results of the Atomistic Simulations

The dislocation was detected by the common neighbor analysis using the Open Visualization Tool (OVITO) [39]. Then a cylinder of height $L$ and radius $R$ was imagined around the dislocation as in Figure 4. The total dislocation energy per unit length $E(R)$ is then given by

$$E(R) = \frac{1}{L}\left(\sum_{r_i \leq R} U_i - NE_{\text{coh}}\right) \quad (4)$$

where $U_i$ is the potential energy of each atom and the summation is over all the atoms inside the cylinder, $N$ is the number of atoms inside the cylinder, and $E_{\text{coh}}$ is the cohesive energy. To keep the external layers of atoms in the imaginary cylinder far from the free surfaces and the replicas of the dislocations, the radius of the cylinder cannot be taken too large.

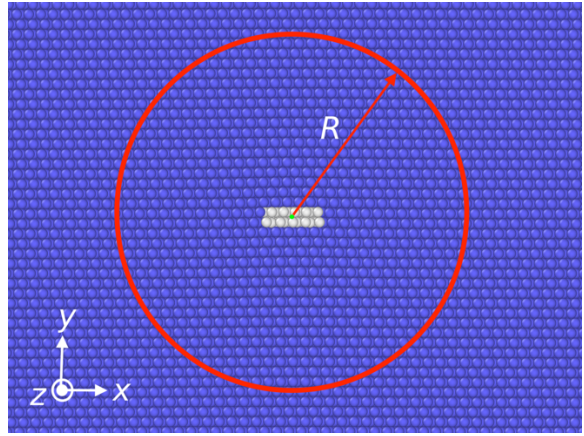

Figure 4: A cylinder of radius *R* and height *L* is considered around the dislocation.

The total dislocation energy is the sum of the elastic energy $E_{\text{el}}$ and the dislocation core energy $E_{\text{core}}$, which is due to non-linear interactions

$$E = E_{\text{core}} + E_{\text{el}}. \tag{5}$$

Elasticity theory predicts that the energy per unit length stored in the medium due to a straight dislocation for $R > r_{\text{core}}$ is linearly proportional to the logarithm of the distance from the dislocation *R*

$$E_{\text{el}} = K_\theta \ln\left(\frac{R}{r_{\text{core}}}\right) = K_\theta \ln R - K_\theta \ln r_{\text{core}}, \tag{6}$$

where $K_\theta$ is the pre-logarithmic energy factor and for isotropic materials is given by [11]

$$K_\theta = \frac{Gb^2}{4\pi}\left(\frac{1}{1-\nu}\sin^2\theta + \cos^2\theta\right). \tag{7}$$

Thus if $R > r_{\text{core}}$

$$E = E_{\text{core}} - K_\theta \ln r_{\text{core}} + K_\theta \ln R \tag{8}$$

Using this fact, the core radius and the core energy are, respectively, the distance and the energy of the point where the strain energy starts to vary logarithmically. The total dislocation energy *E* is plotted versus the distance from the dislocation *R* in Figure 5. The comparison between the slope of *E* versus ln(*R*) for $R \gg r_{\text{core}}$ and $K_\theta$ given by the theory of elasticity shows how well the model is. As we can see from Figure 5, the difference between the slope of the line *E* versus ln(*R*) and the prediction by elasticity is less than 1.5%. Using the anisotropic formula for $K_\theta$ does not make a meaningful difference. We note that because of periodic boundary conditions, there are an infinite number of dislocations, and we need to consider their effects too [40]. However, because the difference between the results of simulations and elasticity theory for a single dislocation is negligible, we can conclude that replicas of the dislocation are so far away that we can ignore them in our calculations.

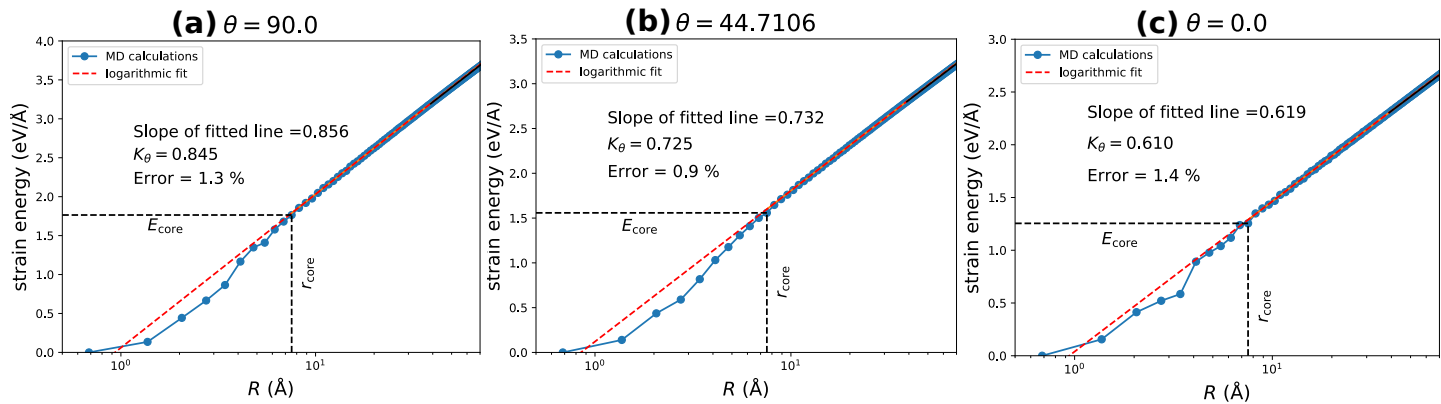

Figure 5: total dislocation energy per unit length as a function of the distance from the dislocation line *R* for three different character angles *θ* (a) when $\theta = 90°$ (edge dislocation) (b) $\theta = 44.71°$ (mixed dislocation) and (c) $\theta = 0$ (screw dislocation). When *R* is greater than the core radius $r_{\text{core}}$, the total energy is a linear function of ln(*R*). The slope of this line is predicted by elasticity. The core energy, $E_{\text{core}}$, is the strain energy when $R = r_{\text{core}}$.

Figure 6 illustrates the total dislocation energy per unit length obtained from the atomistic calculations versus the distance from the dislocation for various character angles. As we can see from Figure 6, $E \propto \ln R$ when $R > 2.75b$. Therefore, we can say that the core radius $r_{\text{core}}$ is about $2.75b$ and $E_{\text{core}} = E(R = r_{\text{core}})$.

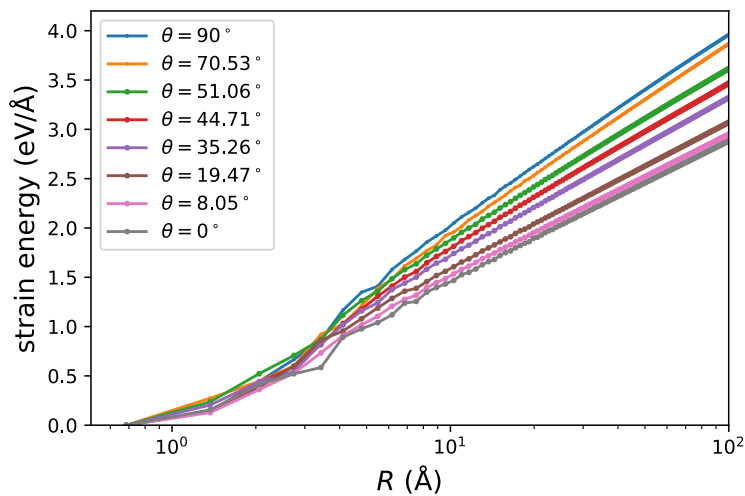

Figure 6: Total dislocation energy versus the distance from the dislocation *r* for various character angles.

If we calculate $E_{\text{core}}$ for each character angle $\theta$ and plot $E_{\text{core}}$ versus $\theta$ (Figure 7), we will realize that we can fit the following curve

$$E_{\text{core}} = E_{\text{core}}^s \cos^2\theta + E_{\text{core}}^e \sin^2\theta, \tag{9}$$

where $E^s_{\text{core}}$ and $E^e_{\text{core}}$ are the core energies of screw and edge dislocations, respectively. For tungsten and for the potential we used, we find

$$E^e_{\text{core}} \approx \frac{1}{1-\nu} E^s_{\text{core}}. \tag{10}$$

Therefore, we can simplify Eq. (9) and rewrite it as

$$E_{\text{core}} \approx E^s_{\text{core}} \left( \cos^2\theta + \frac{1}{1-\nu} \sin^2\theta \right) \tag{11}$$

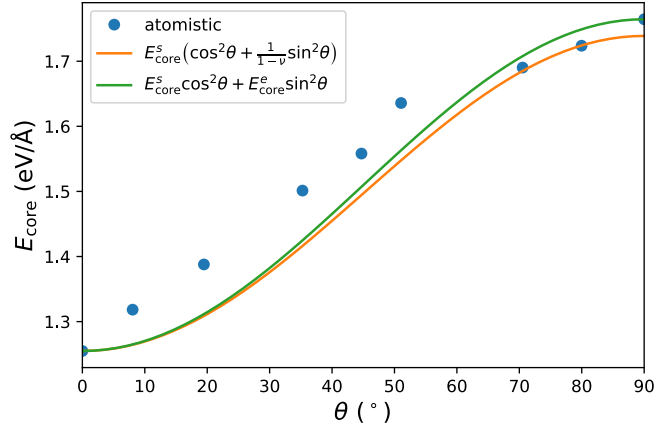

Figure 7: Dislocation core energy $E_{\text{core}}$ obtained from atomistic calculations versus character angle $\theta$. The curves in this figure are the graphs of the function given by Eqs. (9) and (11).

## 5. Classical Elasticity

In the singular dislocation theory of elasticity, the mathematical core radius $r_0$ is arbitrary and is not necessarily taken as $r_{\text{core}}$ (see Section 4) as long as a compensatory core energy $E_c$ is added such that

$$E = E_{\text{core}} + K_\theta \ln\left(\frac{R}{r_{\text{core}}}\right) = E_c + K_\theta \ln\left(\frac{R}{r_0}\right) \tag{12}$$

Therefore, the new core energy $E_c$ differs from $E_{\text{core}}$ that we found in the previous section by $K_\theta \ln(r_0/r_{\text{core}})$:

$$E_c = E_{\text{core}} + K_\theta \ln\left(\frac{r_0}{r_{\text{core}}}\right). \tag{13}$$

Because $E_{\text{core}}$ varies with the character angle $\theta$, $E_c$ also varies with $\theta$. A popular choice for the core radius is $r_0 = b$. For example, if $r_0 = b$, we must take $E_c = 0.938$ eV/Å for an edge dislocation and $E_c = 0.687$ eV/Å for a screw dislocation. Figure 8 represents the variation of $E_c$ with $\theta$ when $r_0 = b$. As we can see from this figure, the equation

$$E_c \approx E_c^s \left(\frac{1}{1-\nu}\sin^2\theta + \cos^2\theta\right), \tag{14}$$

where $E_c^s$ is the core energy for screw dislocation $r_c = b$, fairly describes the dependence of $E_c$ on $\theta$.

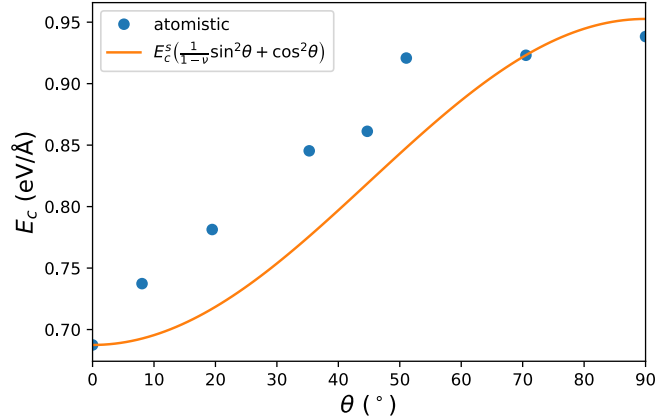

Figure 8: The required core energy when the mathematical core radius is chosen different from the physical core radius $r_{\text{core}}$. In this figure, the cut-off radius is taken as the magnitude of the Burgers vector. The curve is the graph of the function given by Eq. (14).

In some DDD codes, the core energy $E_c$ cannot be specified. Therefore, the core radius has to be taken such that there is no need for the core energy, i.e. $E_c = 0$. In this case, the required core radius, called the effective core radius $r_0^{\text{eff}}$, must be taken as

$$r_0^{\text{eff}} = r_{\text{core}} \exp\left(-\frac{E_{\text{core}}}{K_\theta}\right) \tag{15}$$

This formula is approximate because the slope of the logarithmic fit to the $E$ versus $\ln(R)$ data is slightly different from $K_\theta$ for the reasons explained before.

The effective core radii ($r_0^{\text{eff}}$'s) are calculated for the various character angles (Figure 9). The calculations show us that the variation of the core radius with the character angle is negligible and $r_0^{\text{eff}}$ can be regarded as a constant ($r_0^{\text{eff}} \approx 0.31b$).

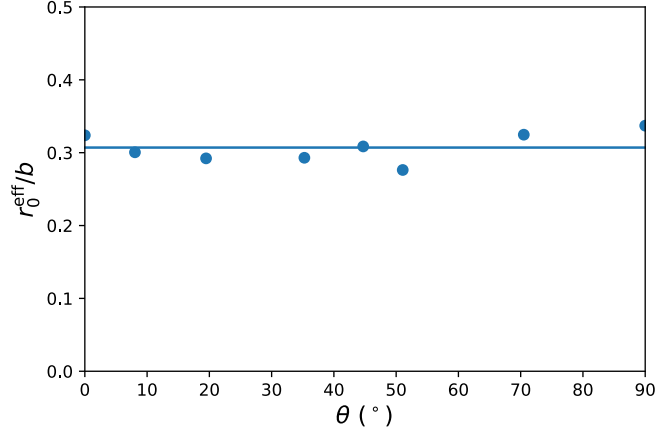

Figure 9: The effective core radius versus the character angle $\theta$

# 6. The CAWB theory

As discussed in Section 1, this theory assumes that the dislocation core is spread out in all three directions according to a distribution function $\tilde{w}(\mathbf{x}; r_c)$ where $r_c$ is a parameter that characterizes the spread of the Burgers vector and is called the core radius [10]. Hence, the stress field evaluated at a given point due to the 'spread-out' dislocation is the original (singular) stress $\sigma_{ij}^0$ convoluted with the distribution function $\tilde{w}(\mathbf{x}; r_c)$. That is, in this theory, the (true) stress at a point, denoted by $\tilde{\sigma}_{ij}$, is

$$\tilde{\sigma}_{ij}(\mathbf{x}) = \sigma_{ij}^0(\mathbf{x}) * \tilde{w}(\mathbf{x}; r_c) = \int \sigma_{ij}^0(\mathbf{x} - \mathbf{x}')\tilde{w}(\mathbf{x}'; r_c)d^3\mathbf{x}' . \tag{16}$$

The elastic energy of a dislocation in this theory then reads

$$E_{\text{el}} = \frac{1}{2}\int S_{ijkl}\tilde{\sigma}_{ij}(\mathbf{x})\tilde{\sigma}_{kl}(\mathbf{x})d^3\mathbf{x} , \tag{17}$$

where $S_{ijkl}$ is the elastic compliance tensor. The issue is that the formula for $\tilde{w}(\mathbf{x}'; r_c)$ and hence the closed form expressions for $\tilde{\sigma}_{ij}$ do not exist. However, the closed form expressions forx the convolution of $\tilde{\sigma}_{ij}(\mathbf{x})$ and $\tilde{w}(\mathbf{x}; r_c)$, which is denoted by $\sigma_{ij}^{\text{ns}}(\mathbf{x})$ and required for calculating the interaction force between two dislocation segments, is available:

$$\sigma_{ij}^{\text{ns}} = \tilde{\sigma}_{ij}(\mathbf{x}) * \tilde{w}(\mathbf{x}; r_c) = \sigma_{ij}^0(\mathbf{x}) * w(\mathbf{x}; r_c), \tag{18}$$

where $w(\mathbf{x}; r_c) = \tilde{w}(\mathbf{x}; r_c) * \tilde{w}(\mathbf{x}; r_c)$. For example, for a screw dislocation

$$\sigma_{\theta z}^{\text{ns}} = \frac{Gb}{2\pi}\frac{\sqrt{x^2+y^2}}{x^2+y^2+r_c^2}\left(1 + \frac{r_c^2}{x^2+y^2+r_c^2}\right). \tag{19}$$

Because $\tilde{w}(\mathbf{x}; r_c)$ can be approximated by

$$\tilde{w}(\mathbf{x}; r_c) \approx 0.3425 w(\mathbf{x}; 0.9038 r_c) + 0.6575 w(\mathbf{x}; 0.5451 r_c), \tag{20}$$

the approximation for $\tilde{\sigma}_{ij}$ is

$$\tilde{\sigma}_{ij}(\mathbf{x}; r_c) \approx 0.3425 \sigma_{ij}^{\text{ns}}(\mathbf{x}; 0.9038 r_c) + 0.6575 \sigma_{ij}^{\text{ns}}(\boldsymbol{x}; 0.5451 r_c). \tag{21}$$

Approximate formulas for the elastic energy per unit length of an edge dislocation $E_{el}^e$ and that of a screw dislocation $E_{el}^s$ when $R \gg r_c$ are [41]

$$E_{el}^e \approx \frac{Gb^2}{4\pi(1-\nu)} \left[ \ln\left(\frac{R}{r_c}\right) + \frac{0.24445 - 0.25\nu - 0.49445\nu^2}{1-\nu^2} \right], \tag{22}$$

and

$$E_{el}^s \approx \frac{Gb^2}{4\pi} \left[ \ln\left(\frac{R}{r_c}\right) + 0.49445 \right]. \tag{23}$$

Again, we can choose the core radius $r_c$ arbitrarily as long as we add an appropriate core energy $E_c$ to compensate for it. That is,

$$E_{\text{core}}(\theta) + K_\theta \ln\left(\frac{R}{r_{\text{core}}}\right) = E_c(\theta; r_c) + E_{el}(R, \theta; r_c). \tag{24}$$

For example, if $r_c = b$, then the core energy for an edge dislocation must be taken to be $E_c^e = E_c(90°; b) = 0.794$ eV/Å and that for a screw dislocation must be taken to be $E_c^s = E_c(0°; b) = 0.386 eV$/Å. The core energy is calculated for various character angle (dots in Figure 10). The following function agrees well with the change of $E_c(\theta; b)$

$$E_c \approx E_c^e \sin^2\theta + E_c^s \cos^2\theta \tag{25}$$

The graph of this function is shown in green in Figure 10. However, the estimation of $E_c^e$ by $E_c^s/(1-\nu)$ does not work well, and as we can see $E_c^s(\frac{1}{1-\nu}\sin^2\theta + \cos^2\theta)$ whose graph is in orange in Figure 10, cannot represent $E_c(\theta; b)$.

These calculations show that the core energy cannot be described by just one number $E_c^s$ and we need at least two extra parameters $E_c^s$ and $E_c^e$ for the description of the core energy.

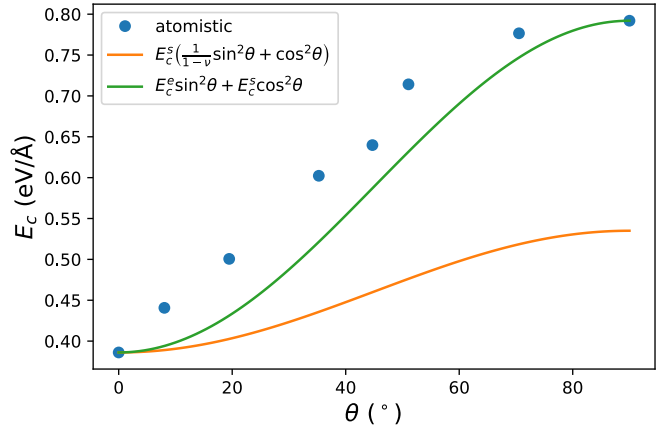

Figure 10: the dots show the required core energy when the core radius in the CAWB theory is taken as the length of the burgers vector. The green curve is the fitted equation given by eq. (25) and the orange curve is the same equation when the core energy for edge dislocation is assumed to be $1/(1-\nu)$ times the core energy of a screw dislocation

If the core energy is not an input for the DDD code, then we have to choose $r_c$ such that the core energy becomes zero (Figure 11). We call this required core radius the effective core radius $r_c^{\text{eff}}$ in the CAWB theory. As we can see from Figure 11, there is no effective core radius that works for every character angle. This means that including the core energy $E_c(\theta)$ in any DDD code that uses the CAWB theory is required.

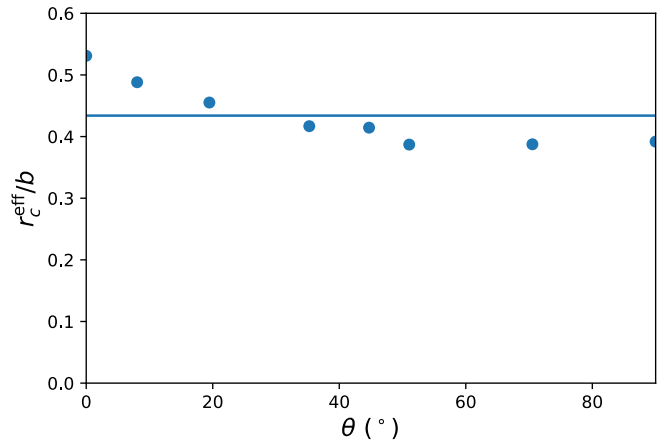

Figure 11: The effective core radius in the CAWB theory for various character angle $\theta$. The horizontal line shows the average value of $r_c^{\text{eff}}$

# 7. Gradient and Nonlocal Theories of Elasticity

Gradient elasticity is a generalization of the classical theory of elasticity and was initially introduced by Mindlin [10–12]. Strain gradient theory includes gradient terms and internal length scales in the constitutive equations to account for microstructures or couple stresses. To remove the singularity of the elastic fields at the dislocation core, Aifantis et al. [20,42,43] introduced a generalized Hooke's law. Specifically

$$(1 - c_1^2 \nabla^2)\boldsymbol{\sigma} = (1 - c_2^2 \nabla^2)(\lambda \, \text{tr}(\boldsymbol{\varepsilon})\mathbf{1} + 2G\boldsymbol{\varepsilon}), \tag{26}$$

where $\boldsymbol{\varepsilon}$ and $\boldsymbol{\sigma}$ denote the elastic strain and stress tensors, $\lambda$ and $G$ are the usual Lamé constants, $\mathbf{1}$ the unit tensor, $\nabla^2$ the Laplacian, $\text{tr}(\boldsymbol{\varepsilon})$ is the trace of the strain tensor, and $c_1$ and $c_2$ are two different gradient coefficients with the dimensions of length. In this theory of gradient elasticity, the stress and strain fields satisfy the following inhomogeneous Helmholtz equations:

$$\begin{aligned}(1 - c_1^2 \nabla^2)\boldsymbol{\sigma} &= \boldsymbol{\sigma}^0, \\ (1 - c_2^2 \nabla^2)\boldsymbol{\varepsilon} &= \boldsymbol{\varepsilon}^0, \end{aligned} \tag{27}$$

where $\boldsymbol{\sigma}^0$ and $\boldsymbol{\varepsilon}^0$ are the singular stress and elastic strain fields given by the classical theory of elasticity. In this theory, if $c_1 = c$ and $c_2 = 0$ or if $c_1 = 0$ and $c_2 = c$, the elastic strain energy for a screw dislocation is [44,45]

$$E_{\text{el}}^s = -\int \sigma_{zy} \varepsilon_{zy}^{*0} dV = \frac{\mu b^2}{4\pi}\left[\ln\left(\frac{R}{2c}\right) + K_0\left(\frac{R}{c}\right) + \gamma\right], \tag{28}$$

where $\varepsilon_{zy}^{*0}$ is the plastic shear strain in the classical theory of elasticity, and $\gamma = 0.5772\ldots$ is the Euler constant. Because $K_0(R/c)$ decays very fast with $R$, if $R \gg c$, the elastic strain energy per unit length reduces to

$$E_{\text{el}}^s = \frac{\mu b^2}{4\pi}\left[\ln\left(\frac{R}{2c}\right) + \gamma\right]. \tag{29}$$

Similarly, the elastic strain energy for an edge dislocation is

$$E_{\text{el}}^e = -\int \sigma_{xy} \varepsilon_{xy}^{*0} dV = \frac{\mu b^2}{4\pi(1-\nu)}\left[\ln\left(\frac{R}{2c}\right) + \gamma - \frac{1}{2} + \frac{2c^2}{R^2} - \frac{2c}{R}K_1\left(\frac{R}{c}\right)\right] \tag{30}$$

and if $R \gg c$, it reads

$$E_{\text{el}}^e = \frac{\mu b^2}{4\pi(1-v)}\left[\ln\left(\frac{R}{2c}\right) + \gamma - \frac{1}{2}\right]. \tag{31}$$

Note that Eqs. (28) and (30) are not singular at the dislocation line where $R \to 0$.

An alternative approach to gradient elasticity is to generalize the strain energy density function and assume that in addition to the classical terms, the strain energy density contains additional terms. Specifically, if

$$E_{\text{el}} = \frac{1}{2}C_{ijkl}\varepsilon_{ij}\varepsilon_{kl} + \frac{1}{2}c^2 C_{ijkl}\partial_p\varepsilon_{ij}\partial_p\varepsilon_{kl}, \tag{32}$$

where $C_{ijkl}$ is the stiffness tensor and $c$ is a length parameter, then the relationship with between classical and gradient fields are also given by Eq. (27) with $c_1 = c_2 = c$ [21,46]. Although in this approach strain field around a dislocation is not singular, the elastic strain energies of screw and edge dislocations are also given by Eqs. (29) and (31), respectively [47].

In nonlocal theory, stress at a point is a function of the strain in the whole body, namely

$$t_{kl} = \int \alpha(|\mathbf{x}' - \mathbf{x}|)\sigma_{kl}^0(\mathbf{x}')d^3\mathbf{x}', \tag{33}$$

where $\alpha(|\mathbf{x}' - \mathbf{x}|)$ is the kernel function and $\sigma_{kl}^0$ is the Hookean stress tensor (or the stress tensor given by the classical theory of elasticity). For two-dimensional problems, the choice of

$$\alpha(|\mathbf{x}|) = \frac{1}{2\pi c^2} K_0\left(\frac{|\mathbf{x}|}{c}\right), \tag{34}$$

where $K_0(.)$ is the modified Bessel function of the second kind of order zero, leads to

$$(1 - c^2 \nabla^2)t_{kl} = \sigma_{kl}^0. \tag{35}$$

That is, in this case, $t_{kl}$ coincides with $\sigma_{kl}$ obtained from Aifantis's gradient elasticity. Because in this theory, displacement and strain remain the same as those in the classical theory of elasticity, the strain energy of a screw dislocation and that of an edge dislocation are given by Eqs. (28) and (30), respectively.

Since in gradient elasticity, often an extra term for the core energy $E_c$ is not included, first we try to see if we can determine a gradient coefficient such that $E_c$ can be neglected. To this end, for each character angle $\theta$, we can find the gradient coefficient $c$ such that

$$E_{\text{el}}^s \cos^2\theta + E_{\text{el}}^e \sin^2\theta = E_{\text{core}}(\theta) + K_\theta \ln\left(\frac{R}{r_{\text{core}}}\right), \tag{36}$$

where $E_{\text{el}}^s$ and $E_{\text{el}}^e$ are given by Eqs. (28) and (30), respectively. For example, when $\theta = 90°$, we find $c = 0.126b$ satisfies Eq. (36) (see Figure 12). We call the solution of Eq. (36) the effective gradient coefficient.

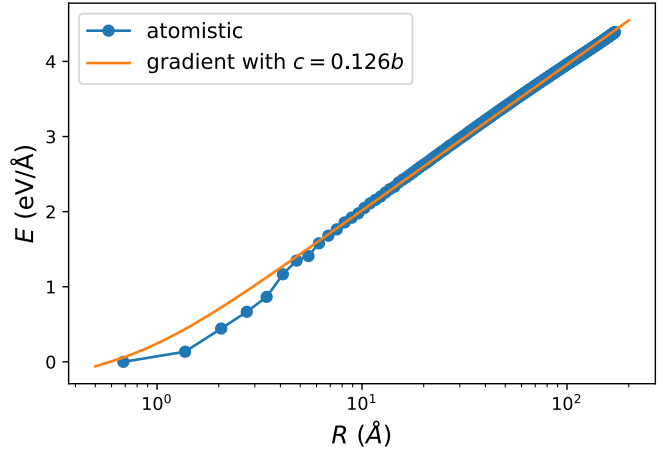

Figure 11: If $c^{\text{eff}} = 0.126b$, the elastic energy per unit length of an edge dislocation in gradient elasticity is very close to the strain energy per unit length obtained from atomistic calculations.

Using Eq. (36), the effective gradient coefficient $c^{\text{eff}}$ is calculated for every character angle (Figure 12). Unlike the effective core radius in classical elasticity, which is almost independent of the character angle, $c^{\text{eff}}$ considerably varies with the character angle $\theta$. For example, for a screw dislocation $\theta = 0$, we have $c^{\text{eff}} = 0.288b$ and for an edge dislocation $\theta = 90°$, we have $c^{\text{eff}} = 0.126b$. We can conclude from Figure 12 that no matter what value for the gradient coefficient is chosen, we have to add a core energy $E_c$.

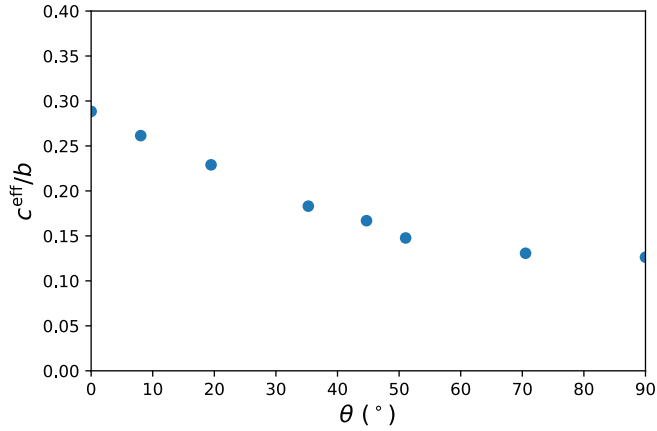

Figure 12: The effective gradient coefficient $c^{\text{eff}}$ in gradient elasticity for various character angles $\theta$

By comparing the stress field around an edge dislocation in tungsten from atomistic simulations and from the theory of gradient elasticity, Po et al. [47] estimated that $c = 0.85b$ gives the best match. Using this value for $c$, $E_c$ is calculated for various character angles (Figure 13). As we can see from this figure, the equation $E_c^s(\frac{1}{1-\nu}\sin^2\theta + \cos^2\theta)$ cannot describe the variation of $E_c$ with the character angle $\theta$, and two parameters $E_c^e$ and $E_c^s$ are required for describing $E_c(\theta)$.

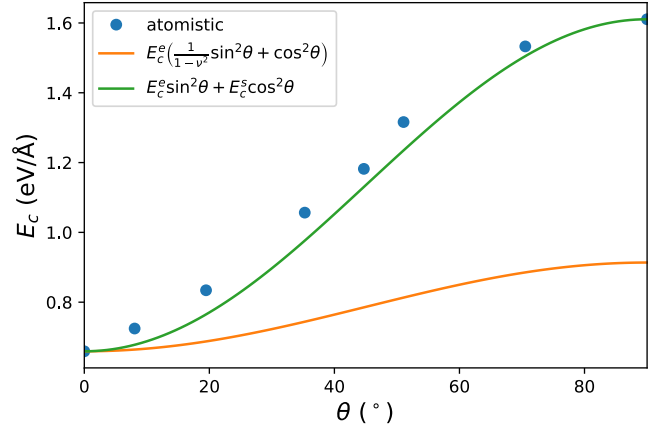

Figure 12: The required core energy per unit length that needs to be added to the elastic energy if $c = 0.85b$.

## 8. Discussion

In this article, we have considered a single dislocation and its infinite replicas. However, as Cai et al [10] suggested, we could consider a dislocation dipole and apply periodic boundary conditions in all three directions. If two dislocations of opposite sign are far away from each other, the energy $E(R)$ approaches a plateau as $R \to \infty$. To show that we would obtain similar results if we used such a configuration, we have considered an edge dislocation dipole with the distance between the two dislocations of the dipole being $d = 200.189$ Å (Figure 14(a)). The strain energy of this configuration as a function of the distance from the center of the dipole $R$ is calculated from the 3atomistic calculations. The results are compared with the predictions by the CAWB theory with the same core radius $r_c$ obtained in Section 6 for a single edge dislocation in Figure 14(b). We can see from Figure 14(b) that the results from these two methods are in good agreement with each other.

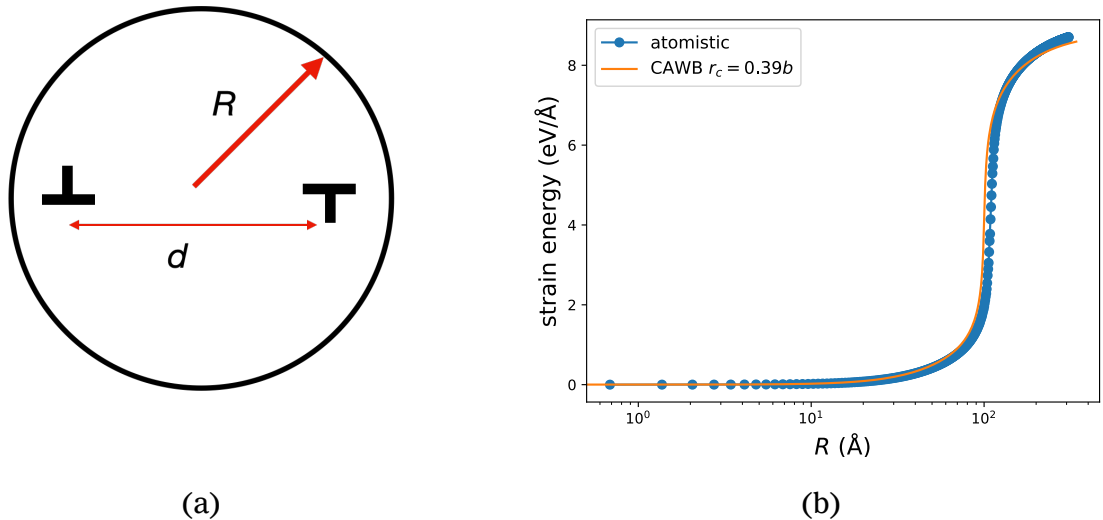

(a)            (b)

Figure 14: (a) an edge dislocation dipole (b) the strain energy of the dipole from atomistic calculations and the prediction by the gradient theory with the effective gradient coefficient $c^{\text{eff}} = 0.39b$.

Because the dependence of $E_{\text{core}}$ on $\theta$ is similar to the dependence of the pre-logarithmic energy factor $K_\theta$ on $\theta$, in Eq. (18) $E_{\text{core}}/K_\theta$ is approximately independent of $\theta$. That is why we can define an effective core radius in classical elasticity that is nearly independent of $\theta$.

By comparing Eqs. (7) and (11), we find that in classical elasticity $E_c(\theta)/K_\theta$ is almost a constant. That is why the effective core radius in classical elasticity is independent of $\theta$

$$r_0^{\text{eff}} \approx r_{\text{core}} \exp\left(-\frac{4\pi E_{\text{core}}^s}{Gb^2}\right).$$

The existence of nonlogarithmic terms in the elastic energy of a dislocation in the CAWB theory (Eqs. 22 and 24) and gradient elasticity (Eqs. 29 and 31) makes $r_c^{\text{eff}}$ and $c^{\text{eff}}$ dependent on $\theta$.

Po et al. [47] determined the gradient coefficient $c$ for tungsten by treating $c$ as a fitting parameter and making the stress components near an edge dislocation as close as possible to the atomistic results. As they have mentioned this method for determining $c$ fails for screw dislocations and perhaps for some mixed dislocation as the virial stress may not present the rise and the inflection point in the theories of elasticity. Additionally, Po et al. [47] employed a theory of gradient elasticity (see Eq. 32) that assumes the stress and the strain at a point are linearly dependent even at the vicinity of the dislocation line [21]. Therefore, after matching the stress components from this theory and from atomistic calculations, strain is also assumed to behave in a specific way, which is not necessarily close to the strain behavior from the atomistic calculations. Consequently, the energy inside in the core region that this theory of gradient elasticity provides might be quite different from the actual core energy $E_{\text{core}}$.

Figure 15 compares the $xy$-component of the stress in classical elasticity, the CAWB theory, and the gradient theory of elasticity in the vicinity of the dislocation line with the effective core cutoff radius, the effective core radius, and the gradient coefficient obtained in the previous sections. From this figure, it is clear that the behavior of stress and the maximum value of stress in these two theories are almost identical provided $r_c$ and $c$ are selected appropriately (i.e. equal to their effective values).

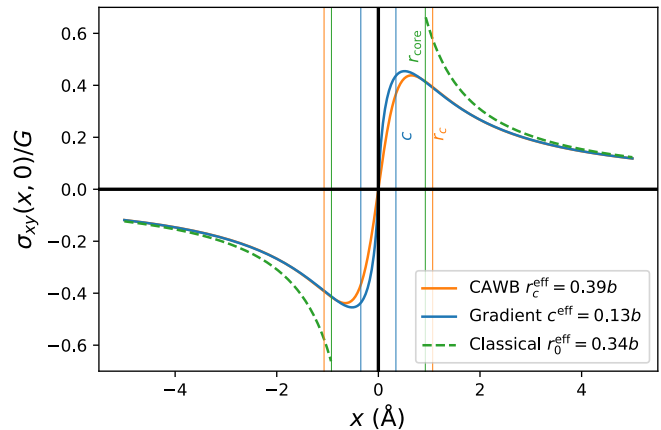

Figure 15: For the effective core radius and the effective gradient coefficient, the stress from the CAWB theory and that from gradient elasticity behave very similarly.

In any of the three theories that we have considered, if the regularization parameter is less than the corresponding effective regularization parameter, the core energy $E_c$ becomes negative. To keep $E_c(\theta)$ positive for every mixed dislocation, $r_0$, $r_c$, and the gradient coefficient $c$ must be larger than the maximum value of $r_0^{\text{eff}}(\theta)$, $r_c^{\text{eff}}(\theta)$, and $c^{\text{eff}}(\theta)$, respectively.

## 9. Conclusions

In this article, we have considered three commonest methods for regularization of the elastic fields in DDD simulations. Each of these three methods contains a regularization parameter with the dimensions of length. We have shown that the values of these regularization parameters significantly affect the simulation results and specifically the stress-strain curve. We have applied atomistic simulations for tungsten to find realistic values for these parameters as well as the radius and core of the physical dislocation core.

Based on the atomistic simulations the physical core radius approximately is $r_{\text{core}} = 2.75b$. The variation of the physical core energy with the dislocation character is similar to the variation of the elastic energy with $\theta$ and can be described only by one parameter: the core energy of a screw dislocation or the that of an edge dislocation.

In classical elasticity if the radius of the mathematical core is taken as $r_0 = b$, then a compensatory core energy which varies with the character angle $\theta$ has to be added. This compensatory energy can be described by one parameter. If the compensatory core region is not included, the core radius has to be taken as $r_0^{\text{eff}} \approx 0.31b$.

To describe the variation of the core energy with the character angle in the CAWB theory and the gradient theory of elasticity, we need the core energy of an edge dislocation and the core energy of a screw dislocation. There is no core radius $r_c$ in the CAWB theory or no gradient coefficient in gradient elasticity that can capture the variation of $E_c$ with $\theta$. When $r_c$ and $c$ are selected such that the core energy for a specific character angle can be ignored, the stresses from both theories behave very similarly.

## 10. Acknowledgement

This research was enabled in part by support provided by the University of Alberta and Compute Canada (www.computecanada.ca). The author would like to thank Dr Ali Tehranchi of Max-Planck-Institut für Eisenforschung GmbH and Prof Wei Cai of Stanford University for helpful discussions.